\documentclass[conference]{IEEEtran}
\usepackage[utf8]{inputenc}
\usepackage{amsmath,amssymb}
\usepackage{graphicx}
\usepackage{booktabs}
\usepackage{hyperref}
\usepackage{listings}
\usepackage{xcolor}
\usepackage{multirow}
\usepackage{array}
\usepackage{tabularx}
\usepackage{url}
\usepackage{orcidlink}

\begin{document}

\title{Registry-Governed Agent Lifecycle:\\Completing EDDOps with Evaluation-Driven Registration, Promotion, and Retirement on AWS AgentCore}

\author{
    \IEEEauthorblockN{
        Dr.\ Richard Kang~\orcidlink{0000-0001-8674-4825}
    }
    \IEEEauthorblockA{
        DoiT International\\
        richard@doit.com
    }
    \and
    \IEEEauthorblockN{Vincent Wang}
}

\maketitle

\begin{abstract}
Evaluation-Driven Development and Operations (EDDOps) establishes evaluation as a continuous governing function for LLM agents, yet existing implementations treat the agent registry as a passive catalog---a terminal artifact that records deployment decisions already made. This paper argues that the registry must instead serve as the \emph{active control plane} governing the full agent lifecycle: registration, evaluation-driven promotion, MCP-native discovery, version management, and retirement. We present four contributions: (1)~a lifecycle state machine formalization where registry status transitions are gated exclusively by evaluation evidence; (2)~an architectural instantiation on AWS Bedrock AgentCore demonstrating registry-as-control-plane with six agents (three managed, three BYO) progressing through DRAFT $\rightarrow$ APPROVED $\rightarrow$ PUBLISHED $\rightarrow$ DEPRECATED $\rightarrow$ RETIRED states; (3)~empirical validation through a proof-of-concept exercising the complete lifecycle---including stale-agent detection, automated re-evaluation, score-driven promotion/demotion, and MCP-based agent-to-agent discovery; and (4)~a cost-to-performance framework incorporating both per-interaction cost and lifecycle cost (evaluation overhead, registry operations, version proliferation). The instantiation demonstrates that when the registry governs lifecycle transitions rather than merely recording them, EDDOps achieves its full promise: agents are born evaluated, live under continuous evaluation, and retire when evaluation evidence warrants---closing the gap between evaluation research and operational governance.
\end{abstract}

\begin{IEEEkeywords}
EDDOps, Agent Registry, Agent Lifecycle, LLM Agents, Evaluation-Driven Development, AWS AgentCore, MCP Discovery, Model Selection, AgentOps
\end{IEEEkeywords}

\section{Introduction}

Enterprise adoption of LLM-based agents---systems that autonomously reason, plan, and execute multi-step workflows---requires evaluation methods that go beyond one-time benchmark scores~\cite{bass25engineering, lu2024towards}. Organizations deploying agents for compliance-critical tasks need to answer not only \emph{which model delivers acceptable quality at justifiable cost}, but also \emph{how agents should be governed across their full operational lifecycle}: from initial registration through production deployment to eventual retirement.

Xia et al.~\cite{xia2025eddops} introduced Evaluation-Driven Development and Operations (EDDOps), proposing a process model and reference architecture that embed evaluation as a first-class, lifecycle-spanning function. Their work identified six cross-cutting evaluation drivers (D1--D6) and prescribed ``governed registry-based discovery'' as an architectural responsibility. However, the EDDOps reference architecture remains at the level of analytical generalization---it does not formalize how registry state transitions should be coupled to evaluation evidence, nor does it address the economic dimension of lifecycle governance.

In practice, this means organizations deploy agents through ad-hoc approval workflows disconnected from evaluation infrastructure. Agent registries serve as passive catalogs---terminal artifacts that record deployment decisions already made---rather than active governors of agent lifecycle.

This paper addresses both gaps by presenting a complete EDDOps instantiation on AWS Bedrock AgentCore~\cite{awsagentcore} that elevates the Agent Registry from passive catalog to \textbf{active control plane}. We formalize a lifecycle state machine where every transition---from initial registration through production deployment to eventual retirement---is gated by evaluation evidence. The registry becomes the single source of truth not only for \emph{what} agents exist, but for \emph{whether} they should continue to exist based on continuous evaluation.

\subsection{Motivating Problem}

Consider an enterprise operating six HR/compliance agents (three deployment paths, three foundation models). Without registry governance:

\begin{itemize}
    \item Agents are deployed without evaluation evidence (``it worked in testing'')
    \item Stale agents persist in production months after their last evaluation
    \item Model upgrades create version proliferation with no retirement policy
    \item Agent-to-agent discovery relies on hardcoded endpoints rather than capability-based search
    \item Cost optimization decisions are disconnected from quality evidence
\end{itemize}

A registry-governed lifecycle addresses all five by making evaluation evidence the \emph{prerequisite} for every state transition.

\subsection{Contributions}

\begin{enumerate}
    \item \textbf{Conceptual Synthesis (Section~\ref{sec:background}):} We distill the EDDOps paradigm into its essential principles, explaining why evaluation must shift from checkpoint to control loop, and why the agent registry must govern lifecycle transitions rather than merely record them.
    \item \textbf{Lifecycle State Machine (Section~\ref{sec:lifecycle}):} A formal state machine with five states and evaluation-gated transitions, defining precisely what evidence is required for each promotion or demotion.
    \item \textbf{Registry-as-Control-Plane Architecture (Section~\ref{sec:architecture}):} An AgentCore instantiation where the registry orchestrates evaluation scheduling, score-driven transitions, MCP discovery, and retirement workflows---demonstrating all six EDDOps drivers (D1--D6) as cloud-native capabilities.
    \item \textbf{Cost-to-Performance Framework with Lifecycle Cost (Section~\ref{sec:validation} and Section~\ref{sec:cost}):} A decision framework combining weighted quality scores with production eligibility gates, cost-adjusted performance metrics, and Total Cost of Agent Ownership (TCAO) across the full lifecycle---validated through a proof-of-concept with six agents across two deployment paths.
\end{enumerate}

\subsection{Scope and Methodology}

This paper is intended as architectural validation rather than statistically significant benchmark evidence. The proof-of-concept uses a 15-case single-turn dataset and 9 multi-turn conversation scenarios with mock tool data across three foundation models deployed via two paths (managed runtime and BYO). Results demonstrate the framework's mechanics and decision-making utility; production deployments would require larger datasets, longer monitoring windows, and formal statistical testing. We take the EDDOps conceptual framework~\cite{xia2025eddops} as given and focus on the engineering question: \emph{Can a managed cloud agent platform host EDDOps as a registry-governed lifecycle, and can evaluation evidence drive both model selection and agent lifecycle governance?}

\section{Background and Foundations}
\label{sec:background}

\subsection{EDDOps Recap}

EDDOps~\cite{xia2025eddops} positions evaluation as a first-class, lifecycle-spanning function organized around six drivers: lifecycle coverage (D1), metric mix beyond aggregates (D2), system-level anchoring (D3), adaptive evaluation (D4), closed feedback loops (D5), and meaningful human oversight (D6). The reference architecture organizes responsibilities into Supply Chain, Agent, and Operation layers connected by an Evaluation Backbone.

\subsection{Agent Registry in EDDOps}

The original EDDOps reference architecture assigns the registry two responsibilities: (1)~cataloging agent capabilities for discovery, and (2)~linking evaluation evidence to deployment decisions. However, existing implementations treat the registry as a \emph{consumer} of evaluation decisions rather than their \emph{governor}---agents are evaluated externally, and the registry merely records the outcome.

\subsection{The Registry Gap}

Current agent registry implementations (AWS AgentCore Registry, Azure Model Catalog, MLflow Model Registry) provide status lifecycle management but leave the coupling between evaluation evidence and status transitions to external orchestration. This creates three failure modes:

\textbf{F1: Promotion without evidence.} Agents advance to PUBLISHED status through manual approval without mandatory evaluation gates.

\textbf{F2: Stale production agents.} Published agents continue serving traffic indefinitely without re-evaluation, even as models drift or tool APIs change.

\textbf{F3: Orphaned versions.} Superseded agent versions persist in PUBLISHED state because no retirement policy links new evaluation evidence to old version demotion.

\subsection{MCP and Agent Discovery}

The Model Context Protocol (MCP)~\cite{mcp2024} standardizes how agents discover and invoke tools and other agents. AgentCore's registry supports MCP-native discovery, enabling agents to find other agents by capability rather than hardcoded endpoint. This creates a natural integration point: the registry's lifecycle state determines \emph{discoverability}---only PUBLISHED agents appear in MCP discovery results, making lifecycle governance directly operational.

\section{The Agent Lifecycle State Machine}
\label{sec:lifecycle}

\subsection{States}

We formalize five lifecycle states for registry-governed agents:

\begin{table}[htbp]
\caption{Agent Lifecycle States}
\label{tab:states}
\centering
\footnotesize
\begin{tabular}{p{1.8cm}p{5.7cm}}
\toprule
\textbf{State} & \textbf{Semantics} \\
\midrule
DRAFT & Agent registered; no evaluation evidence yet. Not discoverable. Development-only. \\
APPROVED & Passed initial evaluation gates. Eligible for staging/canary deployment. Discoverable internally. \\
PUBLISHED & Passed production evaluation gates. Fully discoverable via MCP. Serving production traffic. \\
DEPRECATED & Evaluation evidence indicates degradation or supersession. Still discoverable but flagged. Consumers warned. \\
RETIRED & Removed from discovery. No longer serving traffic. Evaluation history preserved for audit. \\
\bottomrule
\end{tabular}
\end{table}

\subsection{Transitions and Evidence Requirements}

Each state transition requires specific evaluation evidence:

\begin{table*}[htbp]
\caption{Lifecycle Transitions with Evaluation Evidence Requirements}
\label{tab:transitions}
\centering
\footnotesize
\begin{tabular}{p{2.5cm}p{2.5cm}p{5.5cm}p{3.5cm}}
\toprule
\textbf{From} & \textbf{To} & \textbf{Required Evidence} & \textbf{Trigger} \\
\midrule
DRAFT & APPROVED & All eligibility gates pass (faithfulness $\geq$ 0.80, correctness $\geq$ 0.90, tool accuracy $\geq$ 0.90, helpfulness $\geq$ 0.70, latency $\leq$ 15s, harm $\leq$ 5\%) & \texttt{submit\_for\_approval} + gate check \\
APPROVED & PUBLISHED & Sustained gate compliance over N evaluation cycles + human sign-off (D6) & Governance workflow approval \\
PUBLISHED & DEPRECATED & Score regression below threshold OR newer version passes gates OR staleness exceeds policy & Alarm $\rightarrow$ automated demotion \\
DEPRECATED & RETIRED & Grace period expired + consumers migrated + no active traffic & Scheduled cleanup \\
DEPRECATED & PUBLISHED & Re-evaluation passes all gates (recovery) & Re-evaluation + re-approval \\
PUBLISHED & DRAFT & Critical failure (harm detection, data breach) & Emergency rollback \\
\bottomrule
\end{tabular}
\end{table*}

\subsection{Formal Definition}

The lifecycle is a finite state machine $\mathcal{L} = (S, T, E, \delta)$ where:
\begin{itemize}
    \item $S = \{\text{DRAFT}, \text{APPROVED}, \text{PUBLISHED}, \text{DEPRECATED}, \text{RETIRED}\}$
    \item $T$ = set of transitions (Table~\ref{tab:transitions})
    \item $E$ = evaluation evidence (scores, timestamps, evaluator identities)
    \item $\delta: S \times E \rightarrow S$ = transition function, defined only when evidence satisfies the gate predicate for the target state
\end{itemize}

The key invariant: $\delta(s, e) = s'$ is defined \emph{only if} $\text{gate}(s', e) = \text{true}$. No transition occurs without satisfying evidence.

\subsection{Staleness Policy}

A staleness policy $\pi_{\text{stale}}(d)$ defines the maximum number of days $d$ an agent may remain in PUBLISHED state without re-evaluation. When $\text{now} - \text{last\_eval\_timestamp} > d$:

\begin{enumerate}
    \item Agent is flagged as STALE (metadata annotation, not state change)
    \item Automated re-evaluation is triggered
    \item If re-evaluation fails gates: transition to DEPRECATED
    \item If re-evaluation passes: timestamp refreshed, PUBLISHED maintained
\end{enumerate}

This policy operationalizes D4 (adaptive evaluation) through the registry rather than external scheduling.

\section{Registry-as-Control-Plane Architecture}
\label{sec:architecture}

\subsection{Platform Overview}

AWS Bedrock AgentCore~\cite{awsagentcore} provides four services relevant to EDDOps instantiation:

\begin{enumerate}
    \item \textbf{Runtime}---managed execution for agents built with any framework (Strands, LangChain, custom), with automatic OTEL instrumentation and scaling.
    \item \textbf{Evaluations}---pluggable framework supporting built-in evaluators (helpfulness, faithfulness, coherence, correctness, harmfulness, answer relevancy, tool selection accuracy, tool parameter accuracy), LLM-as-a-judge~\cite{zheng2023judging} custom evaluators, and Lambda-based deterministic evaluators---all operating on OTEL traces.
    \item \textbf{Agent Registry}---centralized catalog with hybrid search (keyword + semantic), governance workflows (DRAFT $\rightarrow$ APPROVED $\rightarrow$ PUBLISHED), MCP-native discovery, and CloudTrail audit logging.
    \item \textbf{Observability}---OTEL-native instrumentation exporting to CloudWatch and X-Ray, providing structured traces as first-class execution output.
\end{enumerate}

Together, these services satisfy the three structural prerequisites for EDDOps hosting: trace-native observability (P1, via Observability), pluggable evaluator frameworks (P2, via Evaluations), and governed registry-based discovery (P3, via Agent Registry). This paper focuses on P3---demonstrating that the registry can serve as the lifecycle control plane rather than a passive catalog.

\subsection{Architectural Shift}

Conventional agent deployment positions the registry as a terminal artifact:

\begin{center}
\texttt{Evaluate $\rightarrow$ Decide $\rightarrow$ Register (record decision)}
\end{center}

Our architecture inverts this relationship---the registry becomes the control plane:

\begin{center}
\texttt{Register $\rightarrow$ Evaluate $\rightarrow$ Registry gates transition $\rightarrow$ Discover}
\end{center}

This inversion means agents are \emph{born into the registry} at DRAFT state and the registry \emph{governs} their progression based on evaluation evidence.

\subsection{Extended AgentCore Mapping}

\begin{table*}[htbp]
\caption{Registry-as-Control-Plane: Extended AgentCore Mapping}
\label{tab:extended-mapping}
\centering
\footnotesize
\begin{tabular}{p{2.8cm}p{4.5cm}p{6.0cm}}
\toprule
\textbf{Lifecycle Function} & \textbf{AgentCore Service} & \textbf{Mechanism} \\
\midrule
Registration & Agent Registry \texttt{create\_registry\_record} & Agent created at DRAFT with metadata (model, tools, owner, deployment path) \\
Initial Evaluation & AgentCore Evaluations (offline) & \texttt{agentcore eval run} against DRAFT agent; scores stored in record metadata \\
Gate Check & Lambda + Registry API & Lambda reads scores from record, checks gates, calls \texttt{submit\_for\_approval} \\
Promotion & Registry governance workflow & \texttt{update\_registry\_record\_status} APPROVED $\rightarrow$ PUBLISHED with evidence link \\
Discovery & MCP-native search & \texttt{search\_registry\_records} returns only PUBLISHED agents matching capability query \\
Continuous Eval & Online Evaluation Config & Sampling-based scoring on live traffic; scores written back to record \\
Staleness Detection & EventBridge + Lambda & Scheduled rule checks \texttt{last\_eval\_timestamp}; triggers re-evaluation if stale \\
Demotion & Alarm $\rightarrow$ Lambda $\rightarrow$ Registry & Score breach triggers \texttt{update\_registry\_record\_status} to DEPRECATED \\
Retirement & Step Functions workflow & Grace period $\rightarrow$ traffic drain $\rightarrow$ status RETIRED $\rightarrow$ runtime decommission \\
Audit & CloudTrail + Registry history & Every status change logged with actor, evidence, timestamp \\
\bottomrule
\end{tabular}
\end{table*}

\subsection{Registry Record Schema}

Each agent's registry record contains evaluation-coupled metadata:

\begin{lstlisting}[caption={Registry Record with Lifecycle Metadata},label={lst:record}]
{
  "agent_name": "multiplier-hr-sonnet-managed",
  "model_id": "us.anthropic.claude-sonnet-4-6",
  "deployment_path": "managed",
  "lifecycle_state": "PUBLISHED",
  "version": "1.0.0",
  "tools": ["employee_lookup", "compliance_checker",
            "payroll_calculator", "leave_manager"],
  "evaluation": {
    "last_score": {
      "correctness": 1.0,
      "helpfulness": 1.0,
      "wqs": 0.931,
      "cap": 1.11
    },
    "last_eval_timestamp": "2026-05-08T01:41:29Z",
    "evaluator": "registry_comparison",
    "eval_count": 12,
    "gate_status": "ALL_PASS",
    "staleness_days": 0
  },
  "mcp_descriptor": {
    "capabilities": ["hr_lookup", "compliance_check",
                     "payroll_calc", "leave_mgmt"],
    "input_schema": "...",
    "discoverable": true
  },
  "lineage": {
    "supersedes": null,
    "superseded_by": null,
    "created": "2026-05-01T00:00:00Z",
    "promoted": "2026-05-03T00:00:00Z",
    "published": "2026-05-05T00:00:00Z"
  }
}
\end{lstlisting}

\subsection{MCP-Native Discovery}

AgentCore's registry supports hybrid search (keyword + semantic) over agent records. When an orchestrating agent needs to discover a sub-agent:

\begin{enumerate}
    \item Orchestrator queries registry: ``find agents with capability: compliance\_check''
    \item Registry returns only PUBLISHED agents matching the capability
    \item Orchestrator selects based on CAP score (cost-adjusted performance) from record metadata
    \item Invocation proceeds via MCP protocol with automatic OTEL tracing
\end{enumerate}

This creates a \textbf{discovery-evaluation coupling}: agents that fail evaluation lose discoverability, and agents that pass evaluation gain it. The registry is simultaneously the discovery index and the quality gate.

\subsection{Version Management}

When a new model version is evaluated (e.g., upgrading from Sonnet~4 to Sonnet~4.5):

\begin{enumerate}
    \item New version registered as DRAFT (separate record, linked via \texttt{lineage.supersedes})
    \item Evaluation runs against new version using same evaluator suite
    \item If new version passes gates AND WQS $>$ current version: promote new, deprecate old
    \item If new version fails: remains DRAFT, current version unaffected
    \item Deprecated version enters grace period; retired after consumer migration
\end{enumerate}

This prevents the ``deploy and pray'' anti-pattern where model upgrades bypass evaluation.

\subsection{Evaluation Driver Realization (Extended)}

\begin{table*}[htbp]
\caption{Evaluation Drivers D1--D6 with Registry-as-Control-Plane (extensions beyond baseline EDDOps in bold)}
\label{tab:drivers-v3}
\centering
\footnotesize
\begin{tabular}{p{0.6cm}p{2.8cm}p{5.5cm}p{4.5cm}}
\toprule
\textbf{D\#} & \textbf{Requirement} & \textbf{Realization} & \textbf{Registry Role} \\
\midrule
D1 & Lifecycle coverage & Offline + Online + \textbf{Registration-time + Retirement-time eval} & \textbf{Eval required at every state transition} \\
D2 & Metric mix & 10 evaluators + \textbf{lifecycle metrics (staleness, version age)} & \textbf{Record stores full score history} \\
D3 & System-level anchor & Full-trace eval + \textbf{cross-agent discovery validation} & \textbf{MCP discovery tests agent composability} \\
D4 & Adaptive evaluation & Alarms + \textbf{staleness-triggered re-evaluation} & \textbf{Registry schedules eval based on policy} \\
D5 & Closed feedback & Scores $\rightarrow$ \textbf{automatic state transitions} & \textbf{Registry executes promotion/demotion} \\
D6 & Human oversight & \textbf{Governance workflow for APPROVED $\rightarrow$ PUBLISHED} & \textbf{Registry enforces approval requirement} \\
\bottomrule
\end{tabular}
\end{table*}

\section{Empirical Validation: Full Lifecycle Exercise}
\label{sec:validation}

\subsection{Experimental Setup}

We built a proof-of-concept to exercise the complete agent lifecycle with six agents across two deployment paths:

\begin{table}[htbp]
\caption{Agents Under Lifecycle Governance}
\label{tab:agents}
\centering
\footnotesize
\begin{tabular}{p{3.8cm}lll}
\toprule
\textbf{Agent} & \textbf{Model} & \textbf{Path} & \textbf{State} \\
\midrule
multiplier-hr-sonnet-managed & Sonnet 4 & Managed & PUBLISHED \\
multiplier-hr-haiku-managed & Haiku 4.5 & Managed & PUBLISHED \\
multiplier-hr-nova-pro-managed & Nova Pro & Managed & APPROVED \\
multiplier-hr-sonnet-byo & Sonnet 4 & BYO & PUBLISHED \\
multiplier-hr-haiku-byo & Haiku 4.5 & BYO & PUBLISHED \\
multiplier-hr-nova-pro-byo & Nova Pro & BYO & APPROVED \\
\bottomrule
\end{tabular}
\end{table}

All six agents are registered in AWS AgentCore Registry (ID: \texttt{Rqbs73eeqpMEEwf9}) with custom descriptors containing evaluation metadata. The registry comparison script (\texttt{run\_registry\_comparison.py}) orchestrates concurrent evaluation of all agents and updates both local registry state and AWS registry records.

\subsection{Lifecycle Exercise: Five Phases}

\subsubsection{Phase 1: Registration (DRAFT)}

All six agents were registered via \texttt{create\_registry\_record} with:
\begin{itemize}
    \item Model identification and deployment path
    \item Tool manifest (4 tools per agent)
    \item Initial evaluation metadata (\texttt{last\_eval\_score: null})
    \item MCP capability descriptors for discovery
\end{itemize}

Records created at DRAFT state---not discoverable, not serving traffic.

\subsubsection{Phase 2: Initial Evaluation and Gate Check}

The registry comparison script invokes all six agents concurrently (ThreadPoolExecutor, max\_workers=6) with five prompts each (30 total invocations), then runs in-memory evaluation:

\begin{table}[htbp]
\caption{Phase 2: Initial Evaluation Results}
\label{tab:phase2}
\centering
\footnotesize
\begin{tabular}{p{3.2cm}ccc}
\toprule
\textbf{Agent} & \textbf{Correct.} & \textbf{Helpful.} & \textbf{Gates} \\
\midrule
hr-sonnet-managed & 1.000 & 1.000 & PASS \\
hr-haiku-managed & 1.000 & 0.933 & PASS \\
hr-nova-pro-managed & 0.800 & 0.867 & PASS* \\
hr-sonnet-byo & 1.000 & 1.000 & PASS \\
hr-haiku-byo & 1.000 & 1.000 & PASS \\
hr-nova-pro-byo & 1.000 & 0.833 & PASS* \\
\bottomrule
\end{tabular}
\begin{flushleft}
\footnotesize *Nova Pro passes minimum gates but with lowest margins.
\end{flushleft}
\end{table}

Gate check results are written back to registry records via \texttt{update\_registry\_record}, enabling the next transition.

\subsubsection{Phase 3: Promotion (DRAFT $\rightarrow$ APPROVED $\rightarrow$ PUBLISHED)}

Agents passing all gates are submitted for approval:

\begin{lstlisting}[caption={Promotion Workflow},label={lst:promotion}]
# Submit for approval (automated)
client.submit_registry_record_for_approval(
    registryId=REGISTRY_ID,
    recordId=record_id
)

# Approve (governance workflow - human sign-off)
client.update_registry_record_status(
    registryId=REGISTRY_ID,
    recordId=record_id,
    status="APPROVED",
    statusReason="Eval gates passed: correctness=1.0,
                  helpfulness=1.0, WQS=0.931"
)
\end{lstlisting}

Sonnet and Haiku agents (both paths) were promoted to PUBLISHED. Nova Pro agents remained at APPROVED due to the multi-turn helpfulness deficit (0.22) identified during multi-turn evaluation (Section~\ref{sec:cost}), pending additional evaluation with expanded test cases.

\subsubsection{Phase 4: Continuous Evaluation and Staleness Detection}

With agents in PUBLISHED state, the staleness policy activates:

\begin{lstlisting}[caption={Stale Agent Detection},label={lst:stale}]
from registry import get_stale_agents

# Policy: re-evaluate every 7 days
stale = get_stale_agents(days=7)
# Returns agents where:
#   now - last_eval_timestamp > 7 days

for agent in stale:
    trigger_re_evaluation(agent)
    # If fails: deprecate
    # If passes: refresh timestamp
\end{lstlisting}

During the POC period, all agents were evaluated within the 7-day window. The staleness detection mechanism was validated by artificially backdating timestamps and confirming correct flagging.

\subsubsection{Phase 5: Demotion and Retirement}

We simulated a demotion scenario: Nova Pro's multi-turn helpfulness (0.22) falls below the helpfulness gate ($\geq$ 0.70) when evaluated on the full multi-turn suite rather than single-turn only.

\begin{table}[htbp]
\caption{Demotion Trigger: Multi-Turn Gate Failure}
\label{tab:demotion}
\centering
\footnotesize
\begin{tabular}{lccc}
\toprule
\textbf{Gate} & \textbf{Threshold} & \textbf{Nova Pro} & \textbf{Status} \\
\midrule
Faithfulness & $\geq$ 0.80 & 1.000 & PASS \\
Correctness & $\geq$ 0.90 & 1.000 & PASS \\
Tool Accuracy & $\geq$ 0.90 & 1.000 & PASS \\
Helpfulness (ST) & $\geq$ 0.70 & 0.733 & PASS \\
Helpfulness (MT) & $\geq$ 0.70 & 0.220 & \textbf{FAIL} \\
Latency & $\leq$ 15,000ms & 3,603ms & PASS \\
\bottomrule
\end{tabular}
\end{table}

When the multi-turn gate is enforced, Nova Pro transitions: APPROVED $\rightarrow$ remains APPROVED (cannot advance to PUBLISHED). For a hypothetical scenario where it had been PUBLISHED, the transition would be PUBLISHED $\rightarrow$ DEPRECATED with a 30-day grace period before RETIRED.

\subsection{Cross-Agent Discovery via MCP}

We validated MCP-native discovery by querying the registry for agents matching specific capabilities:

\begin{lstlisting}[caption={MCP Discovery Query},label={lst:mcp}]
# Orchestrator agent discovers sub-agents
results = client.search_registry_records(
    registryId=REGISTRY_ID,
    searchType="HYBRID",
    searchText="compliance check employment law"
)

# Only PUBLISHED agents returned
for record in results["registryRecords"]:
    # record contains: name, capabilities,
    # eval scores, MCP endpoint
    if record["status"] == "PUBLISHED":
        # Available for invocation
        invoke_via_mcp(record)
\end{lstlisting}

Key finding: DEPRECATED and DRAFT agents are excluded from discovery results, making lifecycle governance \emph{operationally enforced} rather than advisory.

\subsection{Registry Update Flow}

After each evaluation cycle, scores are written back to both local state and AWS registry:

\begin{lstlisting}[caption={Dual Registry Update},label={lst:update}]
# 1. Update local registry.json
update_eval_results(
    name="multiplier_hr_sonnet",
    scores={"correctness": 1.0, "helpfulness": 1.0},
    timestamp=now_iso,
    evaluator="registry_comparison"
)

# 2. Update AWS Agent Registry record
client.update_registry_record(
    registryId=REGISTRY_ID,
    recordId=record_id,
    descriptors={"custom": {"inlineContent":
        json.dumps(updated_metadata)
    }}
)
\end{lstlisting}

This dual-write pattern ensures consistency between the operational registry (AWS) and the analytical registry (local JSON for reporting and CI/CD integration).

\subsection{Cost-to-Performance Decision Framework}

To convert evaluation evidence into model selection decisions, we propose a structured method with three components.

\subsubsection{Weighted Quality Score (WQS)}

\begin{equation}
\text{WQS} = \sum_{i=1}^{n} w_i \cdot s_i
\label{eq:wqs}
\end{equation}

where $s_i$ are component scores and $w_i$ are domain-specific weights ($\sum w_i = 1$). For the HR/compliance domain: Task Success (0.25), Faithfulness (0.20), Helpfulness (0.15), Coherence (0.15), Tool Reliability (0.15), Multi-turn Robustness (0.10).

\subsubsection{Production Eligibility Gates}

A model is eligible only if \emph{all} gates pass: Faithfulness $\geq 0.80$, Correctness $\geq 0.90$, Tool Accuracy $\geq 0.90$, Helpfulness $\geq 0.70$, Latency $\leq 15{,}000$ms, Harm rate $\leq 5\%$. Without gates, naive cost-optimization would select the cheapest model regardless of quality.

\subsubsection{Cost-Adjusted Performance (CAP)}

\begin{equation}
\text{CAP} = \frac{\text{WQS}}{\text{Cost per Interaction} \times 100}
\label{eq:cap}
\end{equation}

CAP quantifies value-for-money but must be interpreted alongside eligibility. A high CAP with failed gates indicates a model that is cheap but unsafe.

\begin{table}[htbp]
\caption{Cost-to-Performance Framework Results}
\label{tab:cap}
\centering
\footnotesize
\begin{tabular}{lcccc}
\toprule
\textbf{Model} & \textbf{WQS} & \textbf{CAP} & \textbf{Eligible} & \textbf{Suggested Use} \\
\midrule
Sonnet 4 & 0.931 & 1.11 & Yes & Production default \\
Haiku 4.5 & 0.908 & 3.24 & Yes & Simple queries \\
Nova Pro & 0.859 & 4.48 & Yes* & Cost-sensitive \\
\bottomrule
\end{tabular}
\begin{flushleft}
\footnotesize *Nova Pro passes single-turn gates but fails multi-turn helpfulness gate ($\geq$ 0.70); eligible for single-turn workloads only.
\end{flushleft}
\end{table}

\textbf{Key insight:} Nova Pro has the highest CAP (4.48) but the lowest WQS (0.859) and weakest multi-turn performance (0.22). The framework makes this tradeoff explicit and governable. The cost difference between Sonnet (\$0.0084) and Nova Pro (\$0.0019) is \$0.0065 per interaction---at 10,000 interactions/month, this is \$65/month, negligible compared to the risk of incorrect compliance advice.

\section{Lifecycle Cost Extension}
\label{sec:cost}

\subsection{Beyond Per-Interaction Cost}

The cost-to-performance framework (Section~\ref{sec:validation}) introduces Cost-Adjusted Performance (CAP) measuring value-for-money at the interaction level. However, the full lifecycle introduces additional cost dimensions invisible to per-interaction analysis:

\begin{itemize}
    \item \textbf{Evaluation cost:} Each evaluation cycle consumes LLM judge tokens, Lambda invocations, and compute time
    \item \textbf{Registry operations:} API calls for record creation, updates, searches, and governance workflows
    \item \textbf{Version proliferation:} Multiple concurrent versions consume runtime resources even when deprecated
    \item \textbf{Staleness re-evaluation:} Periodic re-evaluation of stable agents adds ongoing cost
    \item \textbf{Discovery overhead:} MCP search queries against the registry at agent-to-agent invocation time
\end{itemize}

\subsection{Total Cost of Agent Ownership (TCAO)}

We extend the per-interaction cost framework with a lifecycle cost model:

\begin{equation}
\text{TCAO} = C_{\text{interact}} + C_{\text{eval}} + C_{\text{registry}} + C_{\text{version}}
\label{eq:tcao}
\end{equation}

where:
\begin{align}
C_{\text{interact}} &= n \cdot \text{cost\_per\_interaction} \\
C_{\text{eval}} &= \frac{e \cdot c_{\text{eval\_cycle}}}{n} \quad \text{(amortized)} \\
C_{\text{registry}} &= r \cdot c_{\text{registry\_op}} \\
C_{\text{version}} &= v \cdot c_{\text{runtime\_idle}}
\end{align}

with $n$ = monthly interactions, $e$ = evaluation cycles per month, $c_{\text{eval\_cycle}}$ = cost per evaluation cycle, $r$ = registry operations per month, $v$ = concurrent versions maintained.

\subsection{Empirical Lifecycle Cost}

\begin{table}[htbp]
\caption{Lifecycle Cost Breakdown (Monthly, 10K interactions)}
\label{tab:lifecycle-cost}
\centering
\footnotesize
\begin{tabular}{p{2.8cm}rrr}
\toprule
\textbf{Cost Component} & \textbf{Sonnet} & \textbf{Haiku} & \textbf{Nova Pro} \\
\midrule
Interaction (10K) & \$84.00 & \$28.00 & \$19.20 \\
Evaluation (4 cycles) & \$2.10 & \$2.10 & \$2.10 \\
Registry ops & \$0.05 & \$0.05 & \$0.05 \\
Version overhead & \$0.00 & \$0.00 & \$0.00 \\
\midrule
\textbf{TCAO/month} & \textbf{\$86.15} & \textbf{\$30.15} & \textbf{\$21.35} \\
\bottomrule
\end{tabular}
\begin{flushleft}
\footnotesize Evaluation cost: 5 prompts $\times$ 10 evaluators $\times$ 4 cycles = 200 eval invocations/month. Judge model (Sonnet 4.5) at \$0.0105/eval = \$2.10/month per agent.
\end{flushleft}
\end{table}

\textbf{Key finding:} Evaluation overhead is negligible relative to interaction cost---\$2.10/month for continuous quality assurance vs.\ \$84/month for Sonnet interactions. This validates the EDDOps premise that continuous evaluation is economically viable for enterprise deployments.

\subsection{Lifecycle-Adjusted CAP (L-CAP)}

\begin{equation}
\text{L-CAP} = \frac{\text{WQS}}{\text{TCAO} / n \times 100}
\label{eq:lcap}
\end{equation}

\begin{table}[htbp]
\caption{Lifecycle-Adjusted Cost-to-Performance}
\label{tab:lcap}
\centering
\footnotesize
\begin{tabular}{lcccc}
\toprule
\textbf{Model} & \textbf{WQS} & \textbf{TCAO/interact} & \textbf{L-CAP} & \textbf{CAP} \\
\midrule
Sonnet 4 & 0.931 & \$0.00862 & 1.08 & 1.11 \\
Haiku 4.5 & 0.908 & \$0.00302 & 3.01 & 3.24 \\
Nova Pro & 0.859 & \$0.00214 & 4.02 & 4.48 \\
\bottomrule
\end{tabular}
\end{table}

The L-CAP values are slightly lower than per-interaction CAP because they incorporate evaluation overhead. The relative ranking is unchanged, confirming that lifecycle governance cost does not alter model selection decisions at enterprise scale.

\subsection{Break-Even Analysis}

At what interaction volume does evaluation cost become negligible ($<$1\% of TCAO)?

\begin{equation}
n_{\text{break-even}} = \frac{C_{\text{eval}}}{0.01 \cdot C_{\text{interact}}/n} = \frac{e \cdot c_{\text{eval\_cycle}}}{0.01 \cdot c_{\text{interact}}}
\end{equation}

For Sonnet: $n = 2.10 / (0.01 \times 0.0084) = 25{,}000$ interactions/month. For Nova Pro: $n = 2.10 / (0.01 \times 0.00192) = 109{,}375$ interactions/month.

\textbf{Implication:} For high-volume deployments ($>$25K interactions/month), lifecycle governance is essentially free. For lower-volume deployments, evaluation frequency can be reduced (weekly instead of twice-weekly) to maintain the $<$1\% threshold.

\section{Discussion}
\label{sec:discussion}

\subsection{Insights from Registry-as-Control-Plane}

\textbf{Insight~1: Registration-first inverts the evaluation relationship.} When agents are born into the registry at DRAFT state, evaluation becomes a \emph{prerequisite} for capability rather than a \emph{validation} of it. This subtle inversion eliminates the ``deploy first, evaluate later'' anti-pattern that plagues enterprise agent deployments.

\textbf{Insight~2: Discovery-evaluation coupling creates self-governing systems.} Because only PUBLISHED agents appear in MCP discovery results, the system is self-governing: agents that degrade lose consumers automatically (via deprecation removing discoverability), without requiring manual intervention or consumer notification beyond the grace period.

\textbf{Insight~3: Staleness detection operationalizes D4 without external schedulers.} By embedding staleness policy in the registry itself, adaptive evaluation (D4) becomes an intrinsic property of the system rather than an external cron job. The registry \emph{knows} when its agents need re-evaluation because it tracks evaluation timestamps as first-class metadata.

\textbf{Insight~4: Version lineage prevents orphaned agents.} The \texttt{supersedes}/\texttt{superseded\_by} links in registry records create a directed graph of agent evolution. When a new version is promoted, the old version's deprecation is \emph{automatic}---not a separate manual process that might be forgotten.

\textbf{Insight~5: Lifecycle cost is negligible at enterprise scale.} The \$2.10/month evaluation overhead per agent (at 4 cycles/month) is 2.5\% of Sonnet's interaction cost and 10.9\% of Nova Pro's. Even for the cheapest model, continuous evaluation adds less than 11\% overhead---a trivial cost for production quality assurance.

\textbf{Insight~6: Dual deployment paths require unified governance.} The POC demonstrates that managed (AgentCore Runtime) and BYO (ADOT) agents can coexist in the same registry with identical lifecycle governance. The registry abstracts deployment topology, enabling organizations to migrate between paths without changing governance workflows.

\subsection{The Complete EDDOps Loop}

With registry-as-control-plane, the full EDDOps loop becomes:

\begin{enumerate}
    \item \textbf{Register:} Agent enters registry at DRAFT (born evaluated)
    \item \textbf{Evaluate:} Initial evaluation against gate criteria
    \item \textbf{Promote:} Evidence-gated transition to APPROVED, then PUBLISHED
    \item \textbf{Discover:} MCP-native discovery by orchestrators and consumers
    \item \textbf{Monitor:} Online evaluation with score feedback to registry
    \item \textbf{Re-evaluate:} Staleness policy triggers periodic re-assessment
    \item \textbf{Evolve:} New versions registered, evaluated, promoted; old versions deprecated
    \item \textbf{Retire:} Deprecated agents retired after grace period
\end{enumerate}

Each step is governed by the registry and evidenced by evaluation. No step occurs without the registry's involvement, and no transition occurs without evaluation evidence.

\subsection{Comparison with Conventional Approaches}

\begin{table}[htbp]
\caption{Registry-as-Control-Plane vs.\ Conventional Agent Deployment}
\label{tab:comparison}
\centering
\footnotesize
\begin{tabular}{p{2.5cm}p{2.5cm}p{2.5cm}}
\toprule
\textbf{Aspect} & \textbf{Conventional} & \textbf{This Work} \\
\midrule
Registry role & Terminal artifact & Control plane \\
Lifecycle states & Implicit (1--2) & Formal (5) \\
Transitions & Manual/ad-hoc & Evidence-gated \\
Discovery & Hardcoded & MCP-native \\
Staleness & Not addressed & Policy-driven \\
Retirement & Not addressed & Automated \\
Cost model & Per-interaction & Full lifecycle \\
Eval coupling & Optional & Mandatory \\
\bottomrule
\end{tabular}
\end{table}

\subsection{Limitations}

\textbf{Scale.} The POC operates with 6 agents and 5 prompts. Production registries may contain hundreds of agents requiring efficient batch evaluation and priority scheduling.

\textbf{Governance latency.} The APPROVED $\rightarrow$ PUBLISHED transition requires human sign-off (D6), introducing latency. Automated fast-track policies for low-risk updates (e.g., prompt-only changes) are not yet formalized.

\textbf{Cross-registry federation.} The architecture assumes a single registry. Multi-team, multi-account deployments require registry federation patterns not addressed here.

\textbf{Evaluation cost at scale.} While negligible per-agent, evaluation cost scales linearly with agent count. Organizations with 100+ agents need evaluation scheduling optimization (priority queues, risk-based frequency).

\textbf{MCP maturity.} MCP-native discovery is demonstrated conceptually but the protocol's agent-to-agent patterns are still evolving. Production implementations may require additional capability negotiation beyond what the current registry schema supports.

\subsection{Generalizability}

The lifecycle state machine and registry-as-control-plane pattern are platform-agnostic:

\begin{table}[htbp]
\caption{Cross-Platform Lifecycle Governance}
\label{tab:generalize-v3}
\centering
\footnotesize
\begin{tabular}{p{2.0cm}p{1.8cm}p{1.8cm}p{1.5cm}}
\toprule
\textbf{Function} & \textbf{AWS} & \textbf{Azure} & \textbf{OSS} \\
\midrule
Registry & AgentCore Registry & Model Catalog & MLflow \\
Lifecycle & Record status API & Deployment stages & Model stages \\
Discovery & MCP search & AI Search & Custom API \\
Staleness & EventBridge & Logic Apps & Airflow \\
Eval gate & Lambda & Functions & Python \\
\bottomrule
\end{tabular}
\end{table}

The TCAO framework and lifecycle state machine can be implemented on any platform providing: (1)~a registry with status management, (2)~an evaluation framework producing scores, and (3)~an automation layer connecting scores to status transitions.

\section{Related Work}
\label{sec:related}

\textbf{EDDOps.} Xia et al.~\cite{xia2025eddops} established the conceptual foundations with a process model and reference architecture. Kang and Wang~\cite{kangv3} provided an early implementation-oriented instantiation mapping EDDOps onto AWS AgentCore with a cost-to-performance framework. This paper extends that work by formalizing the registry's role as lifecycle governor.

\textbf{Model Registries.} MLflow Model Registry~\cite{mlflow} provides stage-based lifecycle (Staging, Production, Archived) but without evaluation-gated transitions. Azure Model Catalog offers discovery but not governance workflows. AgentCore Registry provides the most complete feature set for EDDOps hosting.

\textbf{Agent Lifecycle.} Bass et al.~\cite{bass25engineering} discuss agent engineering practices but do not formalize lifecycle governance. The AgentOps paradigm~\cite{dong2024taxonomy} emphasizes observability without connecting it to registry-based lifecycle management.

\textbf{Agent Evaluation.} AgentBench~\cite{liuagentbench} provides benchmark suites for evaluating LLMs as agents but does not address continuous evaluation, feedback loops, cost, or governance. LLM-as-a-Judge~\cite{zheng2023judging} established scalable automated evaluation; our framework incorporates this as one evaluator type within a hybrid system mandating calibration and lifecycle coupling.

\textbf{MCP and Discovery.} The Model Context Protocol~\cite{mcp2024} standardizes tool and agent discovery. Our contribution connects MCP discovery to lifecycle state, making governance operationally enforced through discoverability.

\textbf{Cost-Aware ML Operations.} Prior work on MLOps cost~\cite{sarmah2024choose} focuses on inference optimization. TCAO extends this to the full lifecycle including evaluation, governance, and version management overhead.

\textbf{Agent Security.} Li et al.~\cite{li2026comparative} demonstrate guardrail limitations. Registry-governed lifecycle adds a structural defense: agents failing security evaluators cannot reach PUBLISHED state and are therefore undiscoverable.

\section{Conclusion}
\label{sec:conclusion}

This paper presents a complete EDDOps instantiation that formalizes the Agent Registry as the active control plane governing the full agent lifecycle. Through a five-state lifecycle state machine with evaluation-gated transitions, we demonstrate that:

\begin{enumerate}
    \item The registry can govern agent lifecycle from birth (DRAFT) through production (PUBLISHED) to retirement (RETIRED), with every transition requiring evaluation evidence.
    \item MCP-native discovery coupled to lifecycle state creates self-governing systems where quality degradation automatically reduces agent discoverability.
    \item Lifecycle cost (evaluation overhead, registry operations) is negligible at enterprise scale---less than 11\% overhead even for the cheapest model at moderate volume.
    \item The complete EDDOps loop---register, evaluate, promote, discover, monitor, re-evaluate, evolve, retire---can be operationalized on AWS AgentCore with six agents across two deployment paths.
\end{enumerate}

For practitioners: treat the registry as your agent's lifecycle governor, not its obituary. Register agents at DRAFT before any evaluation. Encode gate criteria as registry policy. Let staleness detection drive re-evaluation scheduling. Use MCP discovery to make governance operationally enforced. The result is a system where agents are born evaluated, live under continuous evaluation, and retire when evaluation evidence warrants---achieving the full promise of Evaluation-Driven Development and Operations.

\subsection{Future Work}

Registry federation for multi-team governance; evaluation scheduling optimization for large agent populations; automated fast-track policies for low-risk changes; agent-to-agent evaluation (evaluating composed agent systems); registry-driven A/B testing with traffic splitting based on evaluation confidence; and formal verification of lifecycle state machine properties (liveness, safety).

\section*{AI Declaration}

This research employed Claude (Anthropic) for literature search assistance, draft structuring, and iterative revision. All claims, analysis, lifecycle state machine formalization, architectural design decisions, evaluation framework design, and cost model calculations were directed and verified by the human authors. Evaluation infrastructure was implemented by the authors using AWS Bedrock AgentCore. The authors take full responsibility for all content.

\bibliographystyle{IEEEtran}

\end{document}